\documentstyle[preprint,eqsecnum,aps]{revtex}
\def\bea{\begin{eqnarray}}
\def\eea{\end{eqnarray}}
\def\bes{\begin{eqnarray*}}
\def\ees{\end{eqnarray*}}

\def\simgt{\stackrel{>}{{}_\sim}}
\def\shat{\hat{s}}
\def\that{\hat{t}}
\def\uhat{\hat{u}}
\def\q2{Q^2}
\def\qt{$Q_T$}
\def\pythia{{\tt PYTHIA}}

\def\slashchar#1{\setbox0=\hbox{$#1$}           
   \dimen0=\wd0                                 
   \setbox1=\hbox{/} \dimen1=\wd1               
   \ifdim\dimen0>\dimen1                        
      \rlap{\hbox to \dimen0{\hfil/\hfil}}      
      #1                                        
   \else                                        
      \rlap{\hbox to \dimen1{\hfil$#1$\hfil}}   
      /                                         
   \fi}                                         %
\def\tightenlines{\def\baselinestretch{1.3}\small\normalsize}
\tightenlines
\begin{document}
\draft
\input psfig.sty
\preprint{UCD--99--4}
\title{Higher order corrections to parton\\
showering from resummation calculations}
\author{
S. Mrenna\cite{byline}\\
University of California, Davis\\
Davis, CA 95616}
\date{\today}
\maketitle

\begin{abstract}
The connection between analytic and Monte Carlo calculations of
soft gluon emission is
reanalyzed in light of recent, theoretical developments
in resummation.
An alternative Monte Carlo algorithm is suggested
which incorporates (1) corrections beyond leading order to
the showering and (2) smoothly merges with the higher order
calculation of single, hard parton emission.  
In particular,
it is possible to study jet properties in heavy boson production
for all values of the boson transverse momentum with a
total cross section of NLO accuracy.  The
specific cases of $W$, $Z$ and Higgs boson production at the Tevatron
are addressed using a modified version of the \pythia~Monte Carlo.
\end{abstract}


\narrowtext

\section{Introduction}
\label{sec:intro}

In the near future, experiments at two, high--energy hadron colliders
will search for evidence of physics that supersedes the standard
model.  Important among the tools that will be used in these
searches are showering event generators or showering Monte Carlos
(SMC's).  Among the most versatile and popular of these are the
Monte Carlos {\tt HERWIG} \cite{herwig}, {\tt ISAJET} \cite{isajet}, 
and \pythia \cite{pythia}.  SMC's are useful because they accurately
describe the emission of multiple soft gluons, which is, in effect,
an all orders problem in QCD.  However, they only predict
total cross sections to a leading order accuracy, and, thus,
can demonstrate a sizeable dependence on the choice of scale
used for the parton distribution functions (PDF's) or 
coupling constants (particularly $\alpha_s$).  Also, in general,
they do not translate smoothly into kinematic configurations
where only one, hard parton is emitted.  In distinction
to SMC's are certain analytic calculations which account for
multiple soft gluon emission and higher order corrections to the
hard scattering.  These resummation calculations, however,
integrate out the kinematics of the soft gluons, and, thus,
are limited in their predictive power.  They can, for example,
describe the kinematics of a heavy gauge boson produced in
hadron collision, but cannot predict the number or distribution
of jets that accompany it.  
However, searches for new physics, either directly or indirectly through
careful measurements of standard model predictions, often
demand detailed knowledge of kinematic distributions and
jet activity.
Furthermore, $W$+jets (and $Z$+jets) processes are often backgrounds to
SUSY or technicolor signatures, and we demand a reliable
prediction of their properties.
The aim of this present study is to show how
the positive features of the analytic resummation calculations
can be used to improve the showering algorithms.

The outline of the remainder of this paper is as follows.
First, we review the parton
shower (Sec. II) and the analytic resummation
formalisms (Sec. III).  We then show how to modify the
showing algorithm to incorporate higher--order corrections
to the total cross section (Sec. IV).  Furthermore, we show how to
correct the soft gluon approximation made in the showering, so that
there is a smooth transition between the showering and hard emission
limits (Sec. V),
and we compare this approach to other work.
Finally, we present numerical results (Sec. VI) for $W$, $Z$ and
Higgs boson production at the Tevatron in Run II using
a modified version of \pythia, and our conclusions (Sec. VII).

In the ensuing discussion, we focus on the specific example 
of $W$ boson production at a hadron collider, when the
$W$ decays leptonically.  The results apply equally well
to $\gamma^*, Z$ and Higgs bosons (or any heavy, color--singlet
particle) produced in hadron collisions.

\section{Parton Showers}

SMC's are based on the factorization theorem \cite{factor}, which,
roughly, states that physical observables in any sensible gauge
theory are the product of short--distance functions and long--distance
functions.  The short--distance functions are calculable in
perturbation theory.  The long--distance functions are fit
at a scale, but their evolution to any other scale is also
calculable in perturbation theory.  

A standard application of the factorization theorem is
to describe $W$ boson production at a $p\bar p$ collider
at a fixed order in $\alpha_s$.
The production cross section is obtained by
convoluting the partonic subprocesses evaluated at the
scale $Q$ with the PDF's evaluated at $Q$.  The partons
involved in the hard collision must be sufficiently virtual
to be resolved inside the proton, and a natural choice for
the scale $Q$ is $Q=M_W$ \cite{ISR}.
Prior to the hard collision, however,
the partons are not resolvable in the proton (i.e. the proton
is intact) and have virtualities at a much lower scale $Q_0$ of the order
of 1 GeV.  The connection between the partons at the
low scale $Q_0$ and those at the high scale $Q$ is described by
the DGLAP evolution equations \cite{DGLAP}.  The DGLAP
equations include the most important
kinematic configurations of the splittings $a \to b c$,
where $a,b$ and $c$ represent different types of partons
in the hadron ($q,g$, etc.).  Starting from a measurement
of the PDF's at a low scale $Q_0$, a solution of the DGLAP
equations yields the PDF's at the hard scale $Q$.
Equivalently, starting with a parton $c$ involved
in a hard collision, it is also possible to 
determine probabilistically which splittings generated $c$.
In the process of de--evolving parton $c$ back to the valence
quarks in the proton, a number of 
spectator partons (e.g. parton $b$ in the branching $a\to bc$) 
are resolved.  These partons
constitute a shower of soft jets that accompany the $W$--boson,
and influence its kinematics.  

The shower described above occurs with unit probability and does not change
the total cross section for $W$--boson production calculated
at the scale $Q$ \cite{odorico}.  The showering can be attached to the hard--scattering
process based on a probability distribution {\it after} the
hard scattering has been selected.
Once kinematic cuts are applied, the
transverse momentum and rapidity of the $W$--boson populate
regions never accessed by the differential 
partonic cross section calculated at a fixed order.
This is consistent, since the fixed--order calculation was
inclusive (i.e., $p\bar p\to W+X$) and was never intended to
describe the detailed kinematics of the $W$--boson.  The parton shower,
in effect, resolves the structure of the inclusive state
of partons denoted as $X$.
In practice, the fixed order
partonic cross section (without showering) 
can still be used to describe properties of the decay
leptons as long as the measurable is well defined 
(e.g., the number of leptons with central rapidity 
and high transverse momentum, but not the distribution
of rapidity or transverse momentum of the $W$).  

Here, we review parton showering schematically.
More details can be found in Ref. \cite{kbook}.
First, for simplicity, consider the case of final state or
forward showering, where the parton virtuality  $Q$ evolves
forward to the low scale $Q_0$.  
The basis for developing a probabilistic picture of final state
showering is the DGLAP equation for the fragmentation functions:
\bea
Q {\partial \over \partial Q} D_a(x,Q) =  
\int_{x}^{1-\epsilon}{dz \over z}{\alpha_{abc}(z,Q) \over \pi}
\hat{P}_{a\to bc}(z) D_b(x/z,Q) \nonumber \\
 -D_a(x,Q) \int_{x}^{1-\epsilon}{dz}{\alpha_{abc}(z,Q) \over \pi}
\hat{P}_{a\to bc}(z), 
\eea
where $\hat{P}_{a\to bc}$ is an unregularized splitting function,
$\alpha_{abc}$ is the coupling times color factor, and $\epsilon$ is
a cutoff.  
The equation can be rewritten as
\bea
{\partial \over \partial \ln Q^2} \left( D_a(x,Q)/\Delta(Q) \right) 
 = \int_{x}^{1}{dz \over z}{\alpha_{abc}(z,Q) \over 2\pi}
\hat{P}_{a\to bc}(z) (D_b(x/z,Q)/\Delta(Q)) \nonumber 
\eea
or, after integrating both sides of the expression,
\bea
D_a(x,t') = D_a(x,t) \Delta(t') + \int_{t'}^{t}\int_{x}^{1} dt''
{dz\over z} {\Delta(t') \over \Delta(t'')} {\alpha_{abc}(z,t'') \over 2\pi}
\hat{P}_{a\to bc}(z) D_b(x/z,t''),
\eea
where $t=\ln Q^2,$ with similar definitions for $t'$ and $t''$.
The function 
\bea
\Delta(t') = \exp\left( -\int_{t_0}^{t'}\int_{\epsilon}^{1-\epsilon} dt'' dz {\alpha_{abc}(z,t'') \over 2\pi}
\hat{P}_{a\to bc}(z) \right) 
\eea
gives the probability of evolving from the scale ${Q'}^2=e^{t'}$ to 
$Q_0^2=e^{t_0}$ with
no resolvable branchings, and is called the Sudakov form factor.
$t_0$ is a cutoff scale for the showering.
$\Delta(t')$ contains all the information necessary
to reconstruct a shower, since it encodes the change in virtuality of
a parton until a resolvable showering occurs.
Showering is reduced to iterative solutions
of the equation $r=\Delta(t')/\Delta(t'')$, where $r$ is a random number
uniformly distributed in the interval  $[0,1]$, until a solution for $Q'$ 
is found which is below a cutoff.  For consistency, the cutoff
should represent the lowest scale of resolvable emission $Q_0$.

For the case of initial state radiation, several modifications are necessary.
The fragmentation function is replaced by a parton distribution
function, and the evolution proceeds 
backwards from
a large, negative scale $-|Q^2|$ to a small, negative cutoff scale
$-|Q_0^2|$.  There are two equivalent formulations of backwards
showering based on the probabilities
\bea
\exp\left( -\int_{t'}^{t}\int_{\epsilon}^{1-\epsilon} dt'' dz {\alpha_{abc}(z,t'') \over 2\pi}
\hat{P}_{a\to bc}(z) {x' f_a(x',t')\over x f_b(x,t') } \right), x'=x/z \cite{back1},
\eea
and
\bea
{\Delta(t') \over f_b(x,t')} {f_a(x,t'') \over \Delta(t'') } \cite{back2}.  
\eea 

After choosing the change in virtuality, a particular backwards branching
is selected from the probability function based on
their relative weights (a summation over all possible branchings
$a\to bc$ is implied these expressions), and
the splitting
variable is chosen by solving the equation 
\bea
\int_\epsilon^{x/x'} {dz\over z} \hat{P}_{a\to bc}(z)f(x/z,t')
= r \int_\epsilon^{1-\epsilon} {dz\over z} \hat{P}_{a\to bc}(z)f(x/z,t'),
\eea
where $r$ is a random number.  The details of how a full shower
is reconstructed in the \pythia~Monte Carlo, for example, can be found in Ref.~\cite{pythia}.
The structure of the shower can be complex:
the transverse momentum of the $W$--boson
is built up from the whole series of splittings and boosts, and
is known only at the end of the shower, after the final boost.

The SMC formulation
outlined above is fairly independent of the hard scattering
process considered.  Only the initial choice of partons
and possibly the high scale differs.  Therefore, this
formalism can be applied universally to many different
scattering problems.  In effect, soft gluons are not sensitive to
the specifics of the hard scattering, only the color charge of
the incoming partons.

\section{Analytic Resummation}

At hadron colliders, the partonic cross sections can
receive substantial corrections at higher orders in
$\alpha_s$.  This affects not only the total production
rate, but also the kinematics of the $W$ boson.
At leading order ($\alpha_s^0$), the $W$--boson has a
$\delta(Q_T^2)$ distribution in $Q_T^2$.
At next--to--leading order, the
real emission of a single gluon generates a contribution
to $d\sigma/dQ_T^2$ that behaves as 
$Q_T^{-2}\alpha_s(Q_T^2)$ and $Q_T^{-2}\alpha_s(Q_T^2)\ln(Q^2/Q_T^2)$
while the leading order, soft, and virtual corrections are proportional
to $-\delta(Q_T^2)$.  At higher orders, the most singular terms
follow the pattern of $\alpha_s(Q_T^2)^n\sum_{m=0}^{2n-1}\ln^m(Q^2/Q_T^2)$
$=\alpha_s^n L^m\equiv V^n$.
The logarithms arise from the incomplete cancellation of the virtual
and real QCD corrections, but this cancellation becomes complete for
the integrated spectrum, where the real gluon can become arbitrarily
soft and/or collinear to other partons.  The pattern of singular terms
suggest that perturbation theory should be performed in 
powers of $V^n$ instead of $\alpha_s^n$.
This reorganization of the perturbative series is called resummation.

The first studies of soft gluon emission resummed the leading
logarithms \cite{DDT,Parisi}, leading to a suppression of the cross section
at small $Q_T$.  The suppression
underlies the importance of including sub--leading logarithms \cite{logs}.
The most rigorous approach to the problem of
multiple gluon emission is the Collins--Soper--Sterman
(CSS) formalism for transverse momentum resummation \cite{backtoback},
which resums all of the important logarithms.
This is achieved 
after a Fourier transformation with respect to $Q_T$ in 
the variable $b$, so that the series involving the delta
function and terms $V^n$
simplifies to the form of an exponential.  Hence, the soft gluon
emission is resummed or exponentiated in this $b$--space formalism.  
To be more correct, the Fourier transformation
is the result of expressing the transverse--momentum conserving delta
functions $\delta^{(2)}(\vec{Q}_T - \sum \vec{k}_{T_i})$ in
their Fourier representation.  Also, the exponentiation is
accomplished through the application of the renormalization
group equation, not by reorganizing an infinite sum of terms.
Despite the successes of the $b$--space formalism,
there are several drawbacks.  Most notable for the present study
is that it integrates
out the soft gluon dynamics and does not have a simple
physical interpretation.

The CSS formalism was used by its authors to predict both the
total cross section to NLO and the kinematic distributions
of the $W$--boson to all orders \cite{cssWZ} at
hadron colliders.
A similar treatment was presented using the AEGM formalism \cite{AEGM},
that does not involve a Fourier transform, but is evaluated directly
in transverse momentum $Q_T$ space.  
When evaluated at NLO, the two formalisms are equivalent to NNNL order
in $\alpha_s$, and agree with the fixed order calculation
of the total cross section \cite{AEM85}.  A more detailed numerical comparison of the
two predictions can be found in Ref. \cite{Arnold-Kauffman}.

Recently, the AEGM formalism has been reinvestigated,
and an approximation to the $b$--space formalism has been developed
in $Q_T$--space
which retains its predictive features 
 \cite{resum2} (see also the recent 
eprint \cite{stirling}).
This formulation {\it does} have a simple, physical interpretation,
and can be used to develop an alternate algorithm for 
parton showering which {\it includes} higher--order corrections
to the hard scattering.

In the $b$--space formalism, the differential cross section of the
$W$--boson produced in association with soft gluons is:
\bea
{d\sigma(h_1 h_2 \to V^{(*)}X) \over dQ^2\,dQ^2_T\,dy} =
  {1\over (2 \pi)^2}
  \int_{}^{} d^2 b \, e^{i {\vec b}\cdot{\vec Q_T}}
  \widetilde{W}(b,Q,x_1,x_2) + Y(Q_T,Q,x_1,x_2). 
\eea
In this expression, $Q$, $Q_T$ and $y$ describe the kinematics of the
boson $V$, the function $Y$ is regular as $Q_T\to 0$
and corrects for the soft gluon 
approximation, and the function $\widetilde W$ has the form:
\bea
{\widetilde W} = e^{-S(b,Q)} \Bigg(C \otimes f\Bigg)(x_1,b) 
\Bigg(C \otimes f\Bigg)(x_2,b) H(Q,y),
\eea
where
\bea & &
  S(b,Q,C_1,C_2) =
  \int_{C_1^2/b^2}^{C_2^2Q^2}
  {d {\bar \mu}^2\over {\bar \mu}^2}
       \left[ \ln\left({C_2^2Q^2\over {\bar \mu}^2}\right)
        A\big(\alpha_s({\bar \mu})\big) +
        B\big(\alpha_s({\bar \mu})\big)
       \right],
\eea
and
\bea
  \left( C_{jl} \otimes f_{l/h_1} \right) (x_1,\mu) = 
  \int_{x_1}^{1} {d \xi_1 \over \xi_1} \, 
  C_{jl}( {x_1 \over \xi_1}, C_1, C_2, \mu=C_3/b)
  f_{l/h_1}(\xi_1, \mu=C_3/b).
\eea
In these expressions, $C_1$, $C_2$ and $C_3$ are constants, $H$ is
a function that describes the hard
scattering, and $A$, $B$, and $C$ are calculated perturbatively in
powers of $\alpha_s$:
\bes
(A,B,C)=\sum_{n=0}^{\infty} \left({\alpha_s(\mu)\over\pi}\right)^n (A,B,C)^{(n)}
\ees
(the first non--zero terms in the expansion of $A$ and $B$ are for $n=1$).
The functions $C^{(n)}$ are the Wilson coefficients, and are responsible for
the change in the total production cross section at higher orders.
In fact, $\left( C\otimes f\right)$ is simply a
redefinition of the parton distribution function obtained by
convoluting the standard ones with an ultraviolet--safe function.

We remove the constants $C_1, C_2$ and $C_3$ from these
expressions by choosing their canonical values \cite{backtoback},
which also removes large logarithms.
At leading order, the expression for the 
production of an on--shell $W$--boson simplifies
considerably to:
\bea
 {d\sigma(h_1 h_2 \to WX) \over dQ^2_T} =  \sigma_0
  \int_{}^{} {d^2 b \over (2 \pi)^2} \, e^{i {\vec b}\cdot{\vec Q_T}}
  e^{-S(b,Q)} {f(x_1,b) f(x_2,b) \over f(x_1,Q) f(x_2,Q)},
\eea
\bes
 \sigma_0 =  \kappa \int_{}^{} {dx_1 \over x_1} f(x_1,Q) f(x_2,Q), 
\ees  
where $\kappa$ contains physical constants and we ignore the function $Y$
for now.  The expression contains
two factors, the total cross section at leading order $\sigma_0$,
and a cumulative probability function in $Q_T^2$ that describes the
transverse momentum of the $W$--boson (the total integral over $Q_T^2$
transforms $e^{i\vec{b}\cdot\vec{Q}_T}$ to $\delta^{(2)}(\vec{b})$).
Except for the complication of the Fourier transform, the
term $e^{-S/2} f(x,b)/f(x,Q)$
is analogous to $\Delta(Q) f(x,Q')/\Delta(Q') f(x,Q)$ of the SMC.

Equation (3.1) in $b$--space has a
similar structure in $Q_T$--space.  This is surprising, since
the $b$--space result depends critically on the conservation of total transverse
momentum.  To NNNL accuracy, however, the $Q_T$ space expression agrees
exactly with the $b$--space expression, and has the form \cite{resum2}:
\bea
{d\sigma(h_1 h_2 \to V^{(*)}X) \over dQ^2\,dQ^2_T\,dy} =
  {d\over dQ_T^2}\widetilde{W}(Q_T,Q,x_1,x_2) + Y(Q_T,Q,x_1,x_2). 
\eea
Again ignoring $Y$, we can rewrite this expression as:
\bea
{d\sigma(h_1 h_2 \to WX) \over dQ_T^2 } = \sigma_1
  \left({d\over dQ_T^2}  \left[e^{-S(Q_T,Q)}
  {\left(C\otimes f\right)(x_1,Q_T) \left(C\otimes f\right)(x_2,Q_T) 
   \over \left(C\otimes f\right)(x_1,Q) \left(C\otimes f\right)(x_2,Q)} \right]\right),
\eea
\bes
  \sigma_1 = \kappa \int_{}^{} {dx_1 \over x_1} \left(C\otimes f\right)(x_1,Q) \left(C\otimes f\right)(x_2,Q).
\ees 
The factor $\sigma_1$ is the total cross section to a fixed order, while the rest of 
the function yields the probability that the $W$--boson has a transverse momentum $Q_T$.

\section{A modified showering algorithm}

The primary result of this paper is to exploit the
expression for the differential cross section in Eq.~(3.7), which
has the form of a leading order cross section times a 
backwards evolution.  We 
generalize the function $\Delta(t)/f(x,t)\times f(x,t')/\Delta(t')$ of the
standard backwards showering algorithm
to $\sqrt{\widetilde{W}}$ (the square root
appears because we are considering the evolution of each parton line individually).


To implement this modification in a numerical program, like \pythia, 
we need to provide the new, modified PDF (mPDF) based on the Wilson coefficients.  At leading
order, the only Wilson coefficient is $C_{ij}^{(0)} = \delta_{ij}\delta(1-z)$,
and we reproduce exactly the standard showering formulation.
For $W$--boson production at NLO,
the Wilson coefficients $C$ are:
\bea
C_{jk}^{(1)} = \delta_{jk} \left\{{2\over 3}(1-z) + {1\over 3}(\pi^2-8)\delta(1-z)\right\},
C_{jg}^{(1)} = {1\over 2}z(1-z).
\eea
To NLO, the convolution integrals become:
\bea
\left(C\otimes f_i\right)(x,\mu) = f_i(x,\mu) \left(1+{\alpha_s(\mu)\over\pi} {1\over 3}(\pi^2-8)\right) \nonumber \\
 + {\alpha_s(\mu)\over\pi} \int_{x}^{1} {dz\over z}\left[{2\over 3}(1-z)f_i(x/z,\mu)
 + {1\over 2}(1-z)f_g(x/z,\mu)\right],
\eea
and $f_g(x,\mu)$ is unchanged.
The first term gives the contribution of an unevolved parton to the hard
scattering, while the other two contain contributions from
quarks and gluons with higher momentum fractions that split
$q\to q g$ and $g \to q \bar q$, respectively.  

We are relying on the fact that
the Sudakov form factor used in the analytic expressions and in the SMC are equivalent.  In fact,
the integration over the quark splitting function in $\Delta(Q)$ 
yields an expression similar to the analytic Sudakov:
\bea
\int_{z_m}^{1-z_m} dz C_F \left( {1+z^2\over 1-z} \right) = 
C_F \left( \ln\left[{1-z_m\over z_m}\right]^2
-3/2(1-2z_m) \right) \simeq A^{(1)}\ln(Q^2/Q_T^2) + B^{(1)},
\eea
where $z_m = {Q_T \over (Q+Q_T)}$ is an infrared cutoff, terms of order $z_m$ and higher are neglected, and the $z$ dependence of the
running coupling has been ignored \cite{soft}.
Note that the coefficients $A^{(1)}$ ($C_F$) and $B^{(1)}$ ($-3/2 C_F$)
are universal to 
$q\bar q$ annihilation into a color singlet object, 
just as the showering Sudakov form
factor only knows about the partons and not the details of the hard scattering.
For $gg$ fusion, only the coefficient $A^{(1)}$ (3) is universal.
In general, at higher orders,
the analytic Sudakov is sensitive to the exact hard scattering process.

While the Sudakov form factors are similar, there is no one--to--one
correspondence.  First, the \qt--space Sudakov form factor is expressed
directly in terms of the $Q_T$ of the heavy boson, while, in the SMC's, 
the final $Q_T$ is built
up from a series of branchings.  
Secondly, the integral on the left of Eq.~(4.3) is positive (provided that
$z_m<\frac{1}{2}$), while the analytic expression on the right
can become negative.
This is disturbing, since it means subleading logarithms (proportional
to $B$) are dominating leading ones.  In the exact SMC Sudakov,
the kinematic constraints guarantee that $\Delta(Q)<1$.
In this sense, the Sudakov in the SMC is a more exact implementation
of the analytic one.
We feel that the agreement apparent in Eq. (4.5) 
is compelling enough to
proceed assuming the two Sudakov form factors are equivalent.
In our phenomenological analysis, we calculate $\Delta(Q)$
numerically.

\section{Hard Emission Corrections}

The SMC and resummation formalisms are optimized to deal with kinematic
configurations that have logarithmic enhancements $L$.
For large $Q_T\simeq Q$, there are no such enhancements, and
a fixed order calculation yields the most accurate predictions.
The region of medium $Q_T$, however, is not suited to either
particular expansion, in $\alpha_s^nL^m$ or $\alpha_s^n$.

The problem becomes acute for SMC's.
In the standard implementation of SMC's, the highest $Q_T$ 
is set by the maximum virtuality allowed, $Q=M_W$ in our
example, so that the region $Q_T\ge Q$ is never accessed.
However, at $Q_T \ge Q$, the fixed order calculation is
preferred and yields a non--zero result, 
so there is a discontinuity between the two
predictions.  Clearly, the SMC underestimates the gluon
radiation well before $Q_T \simeq Q$, but the fixed order
calculation make equally overestimate the radiation if extended to
the region $Q_T \le Q$.  

In the $b$--space  calculation, the same sort of behavior would
occur, except that contributions to the cross section that are not
logarithmically enhanced as $Q_T\to 0$ can be added back order--by--order
in $\alpha_s$.  This procedure corrects for the 
approximations made in deriving the exponentiation of soft gluon
emission.  This correction is denoted $Y$,
or $Y_f$ in contrast to the resummed term $Y_r$.  If the coefficients
$A$ and $B$ are calculated to high--enough accuracy, one sees a
relatively smooth transition between Eq.~(3.1) and the NLO prediction
at $Q_T=Q$.  The situation is even better in the $Q_T$--space calculation,
since the matching at $Q_T=Q$ is guaranteed at any order.

It is useful to review  $Y$ in 
the resummed calculation, which has the form
\begin{eqnarray}
Y(Q_T,Q,x_1,x_2)=\int_{x_1}^1{\frac{d\xi _1}{\xi _1}}
\int_{x_2}^1{\frac{d\xi_2}{\xi _2}}\sum_{n=1}^\infty \left[ {\frac{\alpha
_s(Q)}\pi }\right] ^n  f_{a}(\xi _1,Q)\,R_{ab}^{(n)}(Q_T,Q,\frac{x_1}{\xi _1},
\frac{x_2}{\xi _2})\,f_{b}(\xi _2,Q).
\end{eqnarray}
The functions $R$ at first order in $\alpha_s$ are:
\bes
R^{(1)}_{q\bar q} = C_F{ {(\that-\q2)^2 + (\uhat-\q2)^2} \over {\hat{t}\hat{u}} }\delta(\hat{s}+\hat{t}+\hat{u}-Q^2)
 - {1\over Q_T^2}\hat{P}_{q\to q}(z_B)\delta(1-z_A) - (A \leftrightarrow B),
\ees
\bes
R^{(1)}_{gq} = {1\over 2}{ {(\shat+\that)^2 + (\that+\uhat)^2} \over {-\hat{s}\hat{u}} }\delta(\hat{s}+\hat{t}+\hat{u}-Q^2)
- {1\over Q_T^2}\hat{P}_{g\to q}(z_B)\delta(1-z_A),
\ees
and
\bes
R^{(1)}_{qg} = \frac{1}{2}{ (\hat{s}+\hat{u})^2 + (\hat{t}+\hat{u})^2 \over -\hat{s}\hat{t} }\delta(\hat{s}+\hat{t}+\hat{u}-Q^2)
- {1\over Q_T^2}\hat{P}_{g\to q}(z_A)\delta(1-z_B),
\ees
with
\bes
\hat{t}/Q^2 = 1 - 1/z_B\sqrt{1+Q_T^2/Q^2}; \hat{u}/Q^2 = 1 - 1/z_A\sqrt{1+Q_T^2/Q^2},
\ees
\begin{eqnarray}
\hat{P}_{{q\to q}}(z) =C_F\left( {\frac{1+z^2}{{1-z}}}\right),
\hat{P}_{{g\to q}}(z) =\frac{1}{2}\left[ z^2+(1-z)^2\right].
\end{eqnarray}
The first term in each expression is proportional to the 
squared matrix element for the hard emission, while the
terms proportional to $Q_T^{-2}$ are the asymptotic
pieces from the Sudakov form factor.
Similar corrections can be derived for the SMC.
In general, the hard emission contributes to the hadronic cross section proportional to
\bes
   {dx_a dx_b \over \shat} f_{a/h_1}(x_a,Q) f_{b/h_2}(x_b,Q) 
 ~~\sigma_0~~A_{exact},
\ees
where $\sigma_0$ is the leading order partonic cross section,
and the $A_{exact}$ are the expressions multiplying the delta
functions in Eq.~(5.2).
The showering contribution to the same order (assuming showering of only one
parton with virtuality $-\that$) is
\bes 
   dx_a d\eta_b \delta(x_a\eta_b- Q^2/S)f_{a/h_1}(x_a,t) f_{b/h_2}(\eta_b,t) 
~~\sigma_0~~
   {dz \over \that z} \hat{P}(z) {f_{b/h_2}(\eta_b/z,Q) \over f_{b/h_2}(\eta_b,q)}
\Delta(-\that),
\ees
or, by changing variables to $x_b=\eta_b/z$,
\bes
   {dx_a dx_b\over \shat} f_{a/h_1}(x_a,Q) f_{b/h_2}(x_b,Q)~~\sigma_0~~A_{shower}.
\ees
As expressed in the functions $R$ in Eq.~(5.2), 
the single, hard emissions generated by the showering can be
subtracted from the exact squared amplitude to include the remaining NLO corrections
not present in the modified PDF.  This defines a $Y$ term for showering.

The effects of $Y$ are included by generating the NLO
subprocesses and performing the subtraction of showering contributions
on an event--by--event basis.  We illustrate
this explicitly for the $q\bar q'\to W g$ subprocess.  Each event
receives an additional weight $f_Y$ before it is
accepted or rejected (this is accomplished in the subroutine
{\tt PYEVWT} already provided in the \pythia~code for the user
to reweight any process), where $f_Y$ is defined as:
\bes
  f_Y\equiv{{ A_{exact} - A_{shower} }\over{A_{exact}}},
\ees
\bea
  A_{exact} = { {(\that-\q2)^2 + (\uhat-\q2)^2} \over {\hat{t}\hat{u}} },~~~
   A_{shower}= \hat{P}_{q\to q}(z) \left( {\shat \over -\that}\Delta(-\that) +
   {\shat \over -\uhat}\Delta(-\uhat) \right)
\eea
with
$z=Q^2/\shat$.
The correction for $qg\to Wq'$ has a similar form:
\bea
  A_{exact} = { {(\shat+\that)^2 + (\that+\uhat)^2} \over {-\hat{s}\hat{u}} },~~~
   A_{shower}= \hat{P}_{g\to q}(z) {\shat \over -\uhat}\Delta(-\uhat).
\eea
One can show that the showering corrections are either smaller than
the exact squared amplitude or equal to it in the limits
when $\uhat$ or $\that \to 0$.  In writing this expression, we
are ignoring the possibility that later emissions are harder
than the first one \cite{seymour1}.
It has
been shown previously that virtuality--ordered showering, as implemented
in \pythia, yields the hardest emission first 90\% of the time \cite{merging2},
and we ignore this technical detail.

At this point, it is useful to compare the scheme outlined above to
other approaches at improving the showering algorithm.
One class of corrections is 
based on phase--space splitting, where
part of a NLO matrix element is treated with LO kinematics and part
with exclusive NLO kinematics \cite{matching,matching2}.  
The idea is
to allow parton showering for one of these configurations, and
not the other, but particular care must be taken not to mix the
different regions of phase space.
There is an adjustable parameter
that splits the phase space, and physical observables are
sensitive to the exact choice (see the discussion in Ref. \cite{balazs}
regarding $Q_T^{sep}$).  
In the approach of Ref. \cite{matching},
the splitting parameter is tuned so that the contribution with
LO kinematics vanishes.  The resultant showering of the term
with exclusive NLO kinematics generates emissions which are
harder than the first ``hard'' emission, which is not consistent.
Furthermore, the splitting parameters must be retuned for different
processes and different colliders.
This scheme is guaranteed to give
the NLO cross section before cuts, but does not necessarily
generate the correct kinematics.

The other class of corrections modifies
the showering to reproduce the hard emission
limit \cite{merging,merging2}.  While this can be accomplished,
it does so at the expense of transferring events from
low $Q_T$ to high $Q_T$.  There is no attempt to predict
the absolute event rate, but only to generate the correct
event shapes.  In some implementations, the
high scale of the showering is increased to the maximum
virtuality allowed by the collider energy.  This is contrary
to the analytic calculations, where the scale $Q=M_W$, for example,
appears naturally (in the choice of constants $C_1, C_2$ and $C_3$
which eliminate potentially large logarithms).  This scheme
will generate the correct hard limit, but will not generate the
correct cross section in the soft limit.

The present formulation contains the positive features of both schemes.
Phase space is split (but without any adjustable parameters -- which is
also true for the corrections outlined in Ref. \cite{seymour1}), but
higher--order corrections are also applied to the showering algorithm.
This is contained in the modified PDF.
Instead of applying corrections to the showering to reproduce the
hard limit, corrections are applied to the hard emission cross section
to avoid double counting with the showering.  The corrections allow
a smooth transition between the explicit hard emission and the showering.

So far, the discussion has been theoretical.  In the next section, we
demonstrate the phenomenological implementation of these ideas.

\section{Numerical results}

We have applied the showering modifications outlined above to
$W$ and Higgs boson production in Run I and Run II at the
Tevatron, using
a modified version of the \pythia~Monte Carlo \cite{pythia}.
Most of the previous discussion applies exactly to the
case of $Z$ boson production.  In particular, the modified PDF
used for the showering and the corrections $f_Y$ are exactly
the same.  For Higgs boson production, we use the expressions
for the Wilson coefficients presented in Ref. \cite{ABCHiggs}.
Some technical points should be noted.  First, we have not attempted
to modify the Sudakov form factor implemented in \pythia; we have
only modified the PDF's that drive the showering.
The generalization of showering to NLO has only been accomplished for final state showering 
\cite{kato}, and is technically complicated.
Secondly, the expression for $f_Y$, with $z=Q^2/\shat$ and $\Delta(Q)$
calculated numerically using Eq.~(2.3), does not vanish fast enough as
$Q_T\to 0$.  This means that $f_Y$ begins to increase sharply before
vanishing at $Q_T\simeq 0$ because of phase space.  We do not believe this
rise is physical, so we force the vanishing of $f_Y$ below a certain cutoff.
For all cases considered in this study, a 10 GeV cutoff is adequate.
We allow $Y$ to shower with a maximum virtuality fixed at $-\that$ 
or $-\uhat$, and include primordial transverse momentum for
the incoming partons, so this cut is smeared out.  Ideally, however,
no cut would be necessary, and this issue is left for future study.
Finally, we have only considered leptonic decays of
the $W$ and the $\gamma\gamma$ decay of the
Higgs boson (to avoid any effects of final state radiation), 
but our final results are scaled to the total
production cross section.   Our results for $Z$ boson production
are similar to those for $W$ boson production, so we do not
comment on them further.

For our numerical results, 
we present the $Q_T$ distribution of the heavy boson
produced at the Tevatron.  These distributions are in good
agreement with analytic calculations, 
but cannot be predicted accurately by the standard showering algorithm.
Secondly, we present jet properties for the same processes,
which are not significantly altered from the predictions of
the standard showering algorithm.
These cannot be predicted by analytic calculations.

\subsection{Heavy Boson Properties}

Our first goal is to test the predictions of our showering algorithm
on Run I data.  In Fig.~1, the transverse momentum of the $W$ boson (solid line)
is shown
in comparison to D\O~data\cite{d0data} (the three dashed lines show
the data and the upper and lower error estimates).  
The modified PDF (mPDF) is calculated using CTEQ4M PDF's.
As in analytic calculations,
the position of the peak from the SMC depends on non--perturbative physics
\cite{primkt}.
In \pythia, this is implemented through a Gaussian smearing of the
transverse momentum of the incoming partons.  To generate this plot,
we have changed the default Gaussian width from .44 GeV to 3.6 GeV,
which is more in accord with other analyses.  This is the value
used in all subsequent results.  The numerical agreement
between our prediction and the data is very good, except, perhaps, in
the region of 30--50 GeV.\footnote{This agreement also relies on
using \pythia~v6.125 or higher, which treats the showering kinematics
more correctly.  This correction has an even greater impact on Higgs
boson production, where, previously, the transverse momentum distribution
was much broader than for $W$ boson production.}    Our prediction in this region depends on the
details of the correction $f_Y$, and any excess is probably related to the
bad behavior of $f_Y$ at very small $Q_T$.

Next, we present our results on the $Q_T$ distributions for 
$W$ and Higgs bosons ($m_H=100$ GeV) produced at the Tevatron
in Run II.  
Figure~2 displays the $Q_T$ distribution of the $W$ boson at Run II
generated 
using the modified version of \pythia.  The solid line is the full
distribution, and the dashed and dotted lines show the individual contributions
from the corrected showering and the corrected hard emission ($Y$) piece.
For reference, we show the $Q_T$ distribution using the default version
of \pythia~and CTEQ4L PDF's.  Because the maximum virtuality of the
showering is set to the scale $M_W$, the SMC contributions are
suppressed beyond $Q_T\simeq M_W$.  Note how the $Y$ piece fills in
the intermediate $Q_T$ region down to small $Q_T$, where showering
gives the preferred result.  The total cross section
predicted by \pythia~is (18.2, 20.5, 23.9) nb using CTEQ4L,
CTEQ4M, and mPDF$+Y$.  The total increase
in rate from LO to NLO is 
in good agreement with Ref.~\cite{balazs} at
$\sqrt{S}=2$ TeV.  Our numbers for $\sqrt{S}=1.8$ TeV (21.2 nb using mPDF+$Y$)
also agree with the CDF and D\O~data \cite{CDFTotal}.

It is interesting that the mPDF calculation, without $Y$,
yields nearly the same rate as using just the CTEQ4M PDF (21.6 nb vs.
20.5 nb).  This is anticipated by the smallness of the virtual correction
in the Wilson coefficients $\propto \pi^2-8$.  However, this need not
be the case for different processes or different colliders.

The results for the production of a Higgs boson with mass $m_H=100$ GeV is
shown in Fig.~3.  Here, the correction to the lowest order process is quite
large.  The total cross section
predicted by \pythia~is (0.50, 0.48, 0.94) pb using CTEQ4L,
CTEQ4M, and the mPDF$+Y$.  The final number are in good agreement
with Ref.~\cite{TeVHiggs}.
We have used a 
primordial $k_T$ tuned to Run I data.
However, if we believe that the non--perturbative function
should have the same form as the perturbative Sudakov form factor, then 
the primordial $k_T$ should scale like $C_A/C_F$ relative to the $q\bar q$ case \cite{renormalon},
and the peak of the $Q_T$ distribution would shift to the right.

It is interesting to know if the kinematic distributions for the heavy bosons can 
be reproduced using the standard showering algorithms with a multiplicative
$K$--factor that yields the total NLO rate.   Figure~4 shows the ratio of
the Higgs boson transverse momentum distributions calculated from mPDF+$Y$
to CTEQ4L times $K$ ($K\simgt 2$).  There are variations as large as 10\%
in the important regions of small and medium $Q_T$.   Of course, the effect
is much larger for the large $Q_T$ region where there is almost no rate
from the standard parton showering.

\subsection{Jet properties}
In Figs.~5, 6 and 7, we present jet properties
for the $W$ and Higgs boson production processes.
Jets are defined using the cone clustering algorithm of
the \pythia~subroutine {\tt PYCELL} with a cone of size
$R=0.5$, $E_T>5$ GeV, and $|\eta|<2.5$.  For $W$ boson
production in Fig.~5, we present the jet multiplicity distributions
for all \qt, $Q_T<10$ GeV, $10<Q_T<20$ GeV, and $Q_T<30$ GeV.
We compare the showering predictions using mPDF (solid line),
CTEQ4M (dashed line), and CTEQ4L (dotted line).  To study
the effects of the modified showering algorithm, we do not
include $Y$, which would increase the $N_{jet}=1$ bin of the
mPDF prediction.  All distributions are normalized to unity.
From Fig.~5, we see that the predictions have only minor 
differences, which is expected since the Wilson coefficients
for $W$ production are nearly unity.

For the case of Higgs boson production in Fig.~6, we study
the regions $Q_T<25$ GeV, $25<Q_T<50$ GeV, and $Q_T<75$ GeV.
There are more noticeable differences, and much 
more radiation in general than for the $W$ boson case.
The higher order PDF's generate slightly more jet activity
and yield similar distributions.  In general, though, there
are no dramatic changes in the distributions.  This is not
too surprising, since the showering depends on the ratio
of the modified PDF's evaluated at two different scales,
which is not as sensitive to the overall normalization
of the PDF.

From these examples, we learn that the use of the modified
showering algorithm does not change the jet properties in
a major way.  Figure 7 compares the differential jet
transverse energy distribution in Run I to CDF data.  Here,
jets are defined with $R=0.4$ and are smeared with
an energy resolution function $\Delta E_T/E_T = 1.2/\sqrt{E_T}$ (in GeV).
We consider the agreement between theory and data to be 
further evidence that our NLO corrections reproduce 
jet properties.

\section{Conclusions}

We have presented a modified, parton showering algorithm that 
produces the total cross section and the event shapes beyond
the leading order.
These modifications are based on the $Q_T$--space
resummation.  The parton showering itself is modified by using
a new PDF (called mPDF) which encodes some information about the hard
scattering process.  Simultaneously, the explicit, hard emission
is included, but only after subtracting out the contribution
already generated by the showering: this correction
is called $Y$.  The presence of $Y$ yields  a smooth
transition from the parton showering to single, hard emission.   
We modified the
\pythia~Monte Carlo to account for these corrections, and
presented comparisons with Run I $W$ boson data and predictions
for $W$ and Higgs boson production
at the Tevatron in Run II.  

The scheme works very well at NLO for the cases considered in
this study, and the correct cross sections, transverse momentum
distributions, and jet properties are generated.  
We have compared our kinematic distributions to the case when
the results of the standard showering are multiplied by a constant
$K$--factor to reproduce the NLO cross section.  We find variations
on the order of 10\% for small and medium transverse momentum.

There are several effects which still need study.
We have not included the exact distributions for the decay
of the leptons \cite{Balazs-Qui-Yuan} for $W$ and $Z$ production,
which are resummed differently.
It is straightforward to include such effects.
In the theoretical discussion and numerical results, we have focussed
on initial state radiation, 
but our results should apply equally well
for final state radiation.  The situation is certainly simpler, since final
state radiation does not require detailed knowledge of the
fragmentation functions.  Also, the case when color flows from
the initial state to the final state requires study.
A resummed calculation already exists for the case of
deep inelastic scattering \cite{Olness}, and
much theoretical progress has been made for heavy quark
production \cite{Sterman}.
We believe that the modified showering scheme outlined in
this study
generalizes beyond NLO, just as the analytic calculations
can be calculated to any given order.
For example, we could include hard $W+2$ jet corrections \cite{Arnold-Reno}
to $Y$.  For consistency, however, higher order
terms ($A$ and $B$)  may
also need to be included in the Sudakov form factor.

The modified \pythia~subroutines used in this study
and an explanation of how to use them
are available at the web address moose.ucdavis.edu/mrenna/shower.html.

\acknowledgements
We thank C--P Yuan and T. Sj\"ostrand for many useful discussions and encouragement
in completing this work.  This work was supported by United States Department of
Energy and the Davis Institute for High Energy Physics.

%
%
\begin{center}
\begin{figure}
\psfig{figure=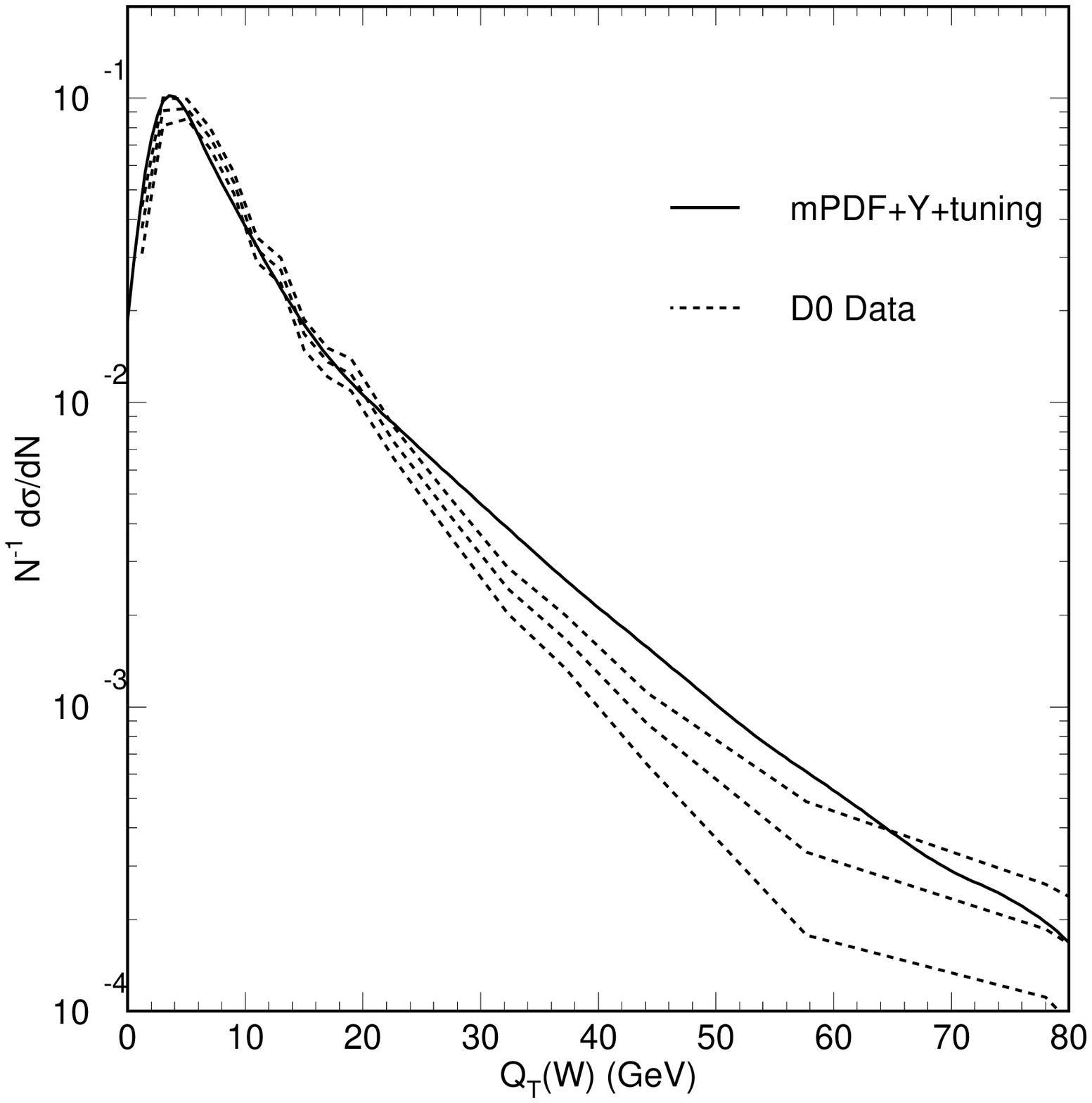,width=15cm}
\caption[]
{The prediction of
the $W$ boson transverse momentum distribution in Run I at the
Tevatron (solid line) compared to the D\O~data (the upper and
lower dashed lines represent the errors on the middle line).
The prediction includes the effects of the modified parton
distribution functions, the correction to the hard scattering
process, and a tuned primordial $k_T$ of 3.6 GeV.}
\label{fig:wpt_runi}
\end{figure}
\end{center}
%
\begin{center}
\begin{figure}
\psfig{figure=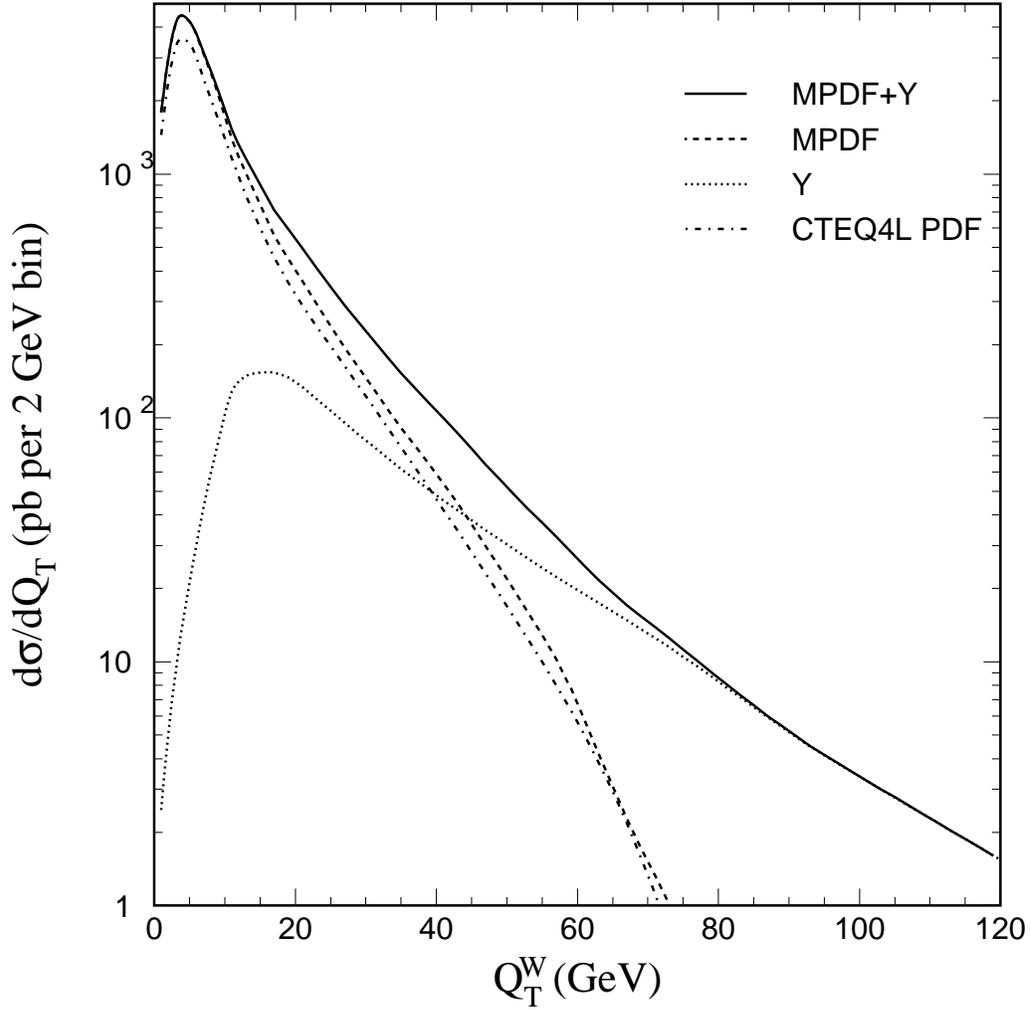,width=15cm}
\caption[]
{Predicted $W$ boson transverse momentum distribution in
Run II (solid line) showing the individual contributions
from showering (long dashes), the corrected hard emission
(short dashes) and the standard \pythia~prediction using
CTEQ4L structure functions (dot--dash).}
\label{fig:wpt}
\end{figure}
\end{center}

\begin{center}
\begin{figure}
\psfig{figure=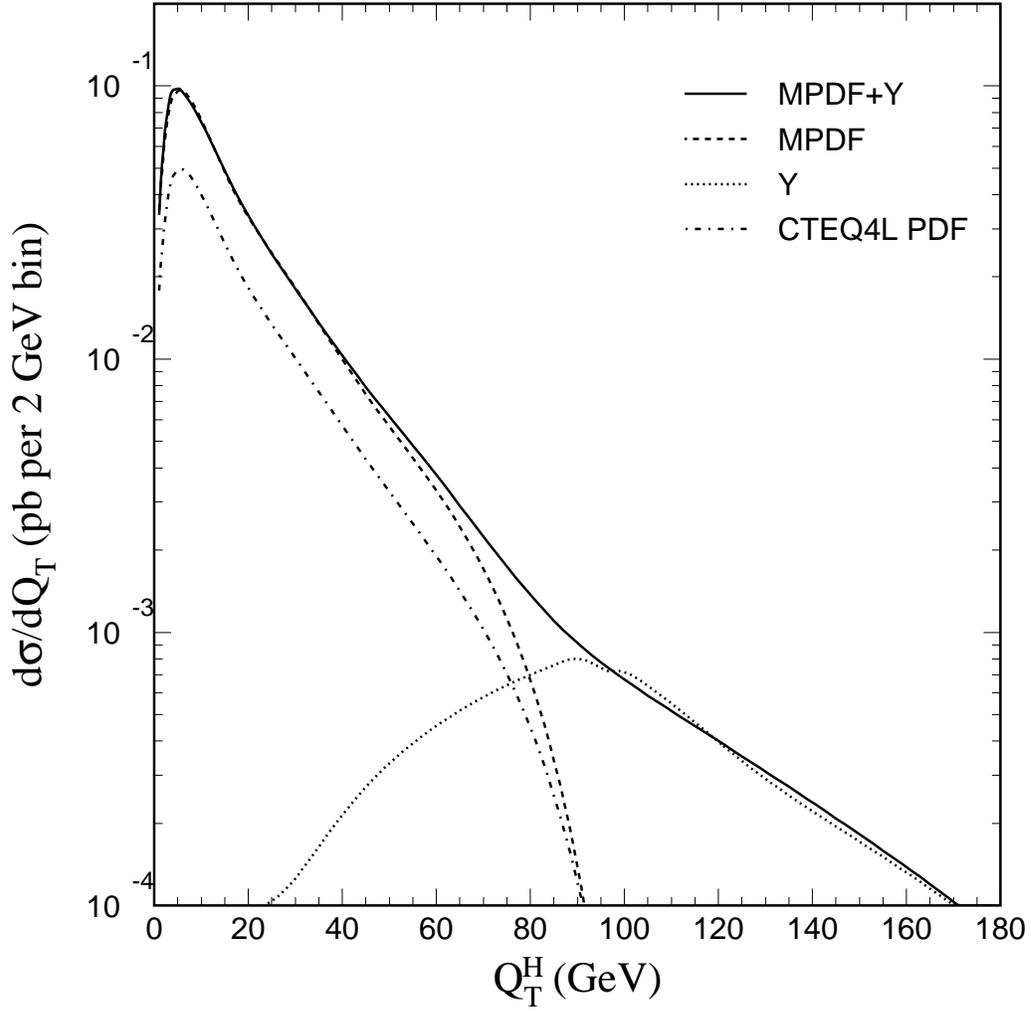,width=15cm}
\caption[]
{Same as Fig.~2, but for Higgs boson production ($m_H=100$ GeV).}
\label{fig:hpt}
\end{figure}
\end{center}

\begin{center}
\begin{figure}
\psfig{figure=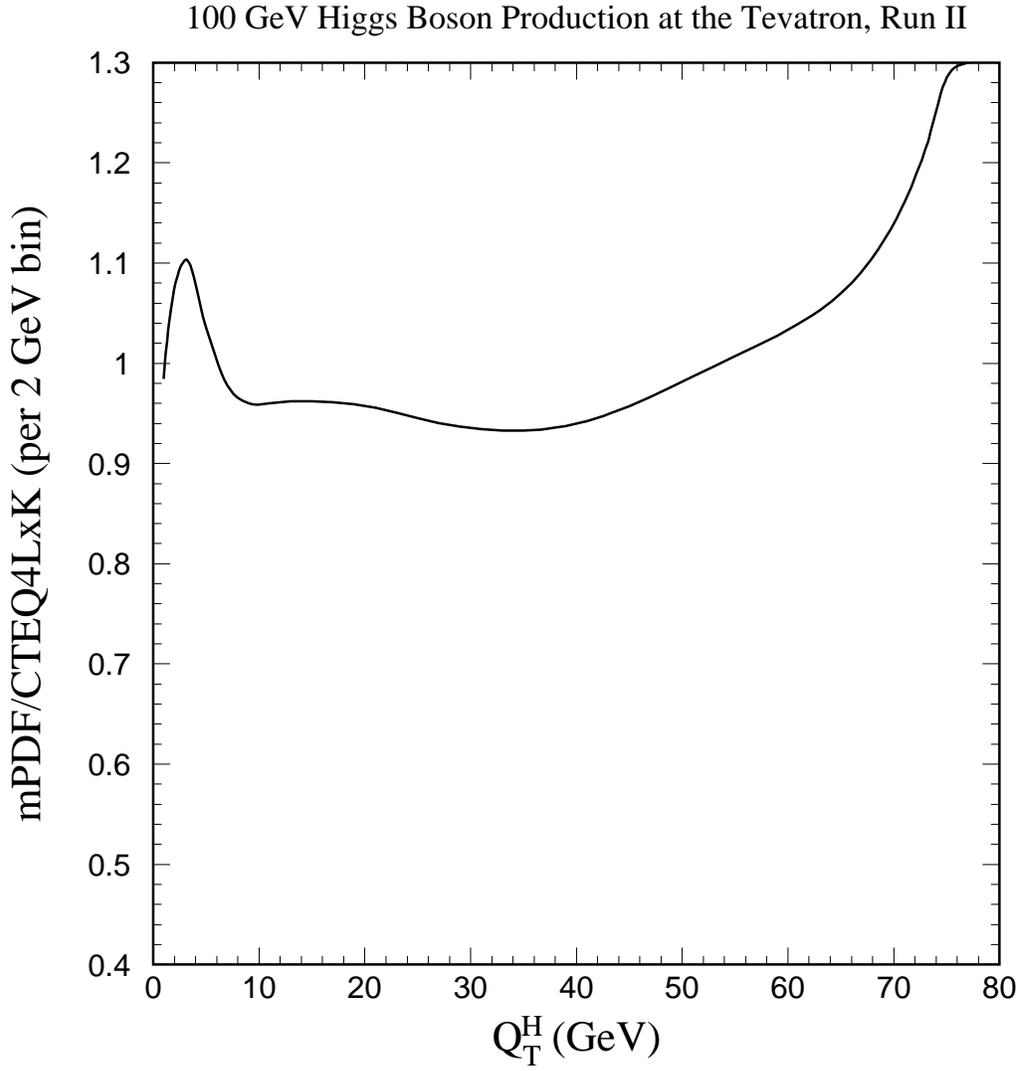,width=15cm}
\caption[]
{The ratio of Higgs boson transverse momentum distributions ($m_H=100$ GeV)
from the modified showering algorithm and from the standard showering algorithm
using the CTEQ4L PDF.  The CTEQ4L result has been multiplied by a constant 
$K$--factor to reproduce the NLO rate.}
\label{fig:ratio}
\end{figure}
\end{center}

\begin{center}
\begin{figure}
\psfig{figure=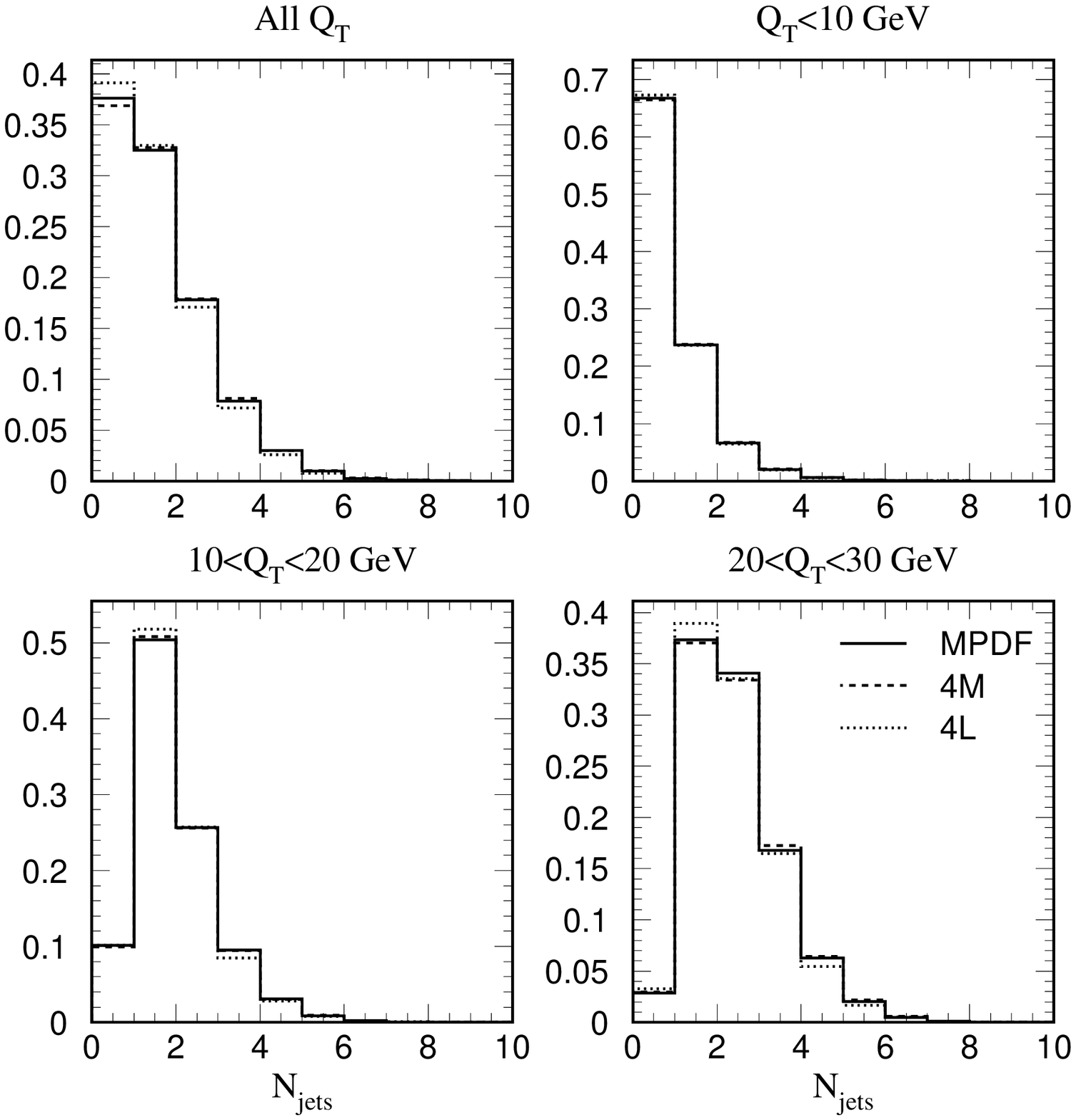,width=15cm}
\caption[]
{Jet multiplicity in $W$ boson events for different $Q_T$ binnings.}
\label{fig:wjets}
\end{figure}
\end{center}

\begin{center}
\begin{figure}
\psfig{figure=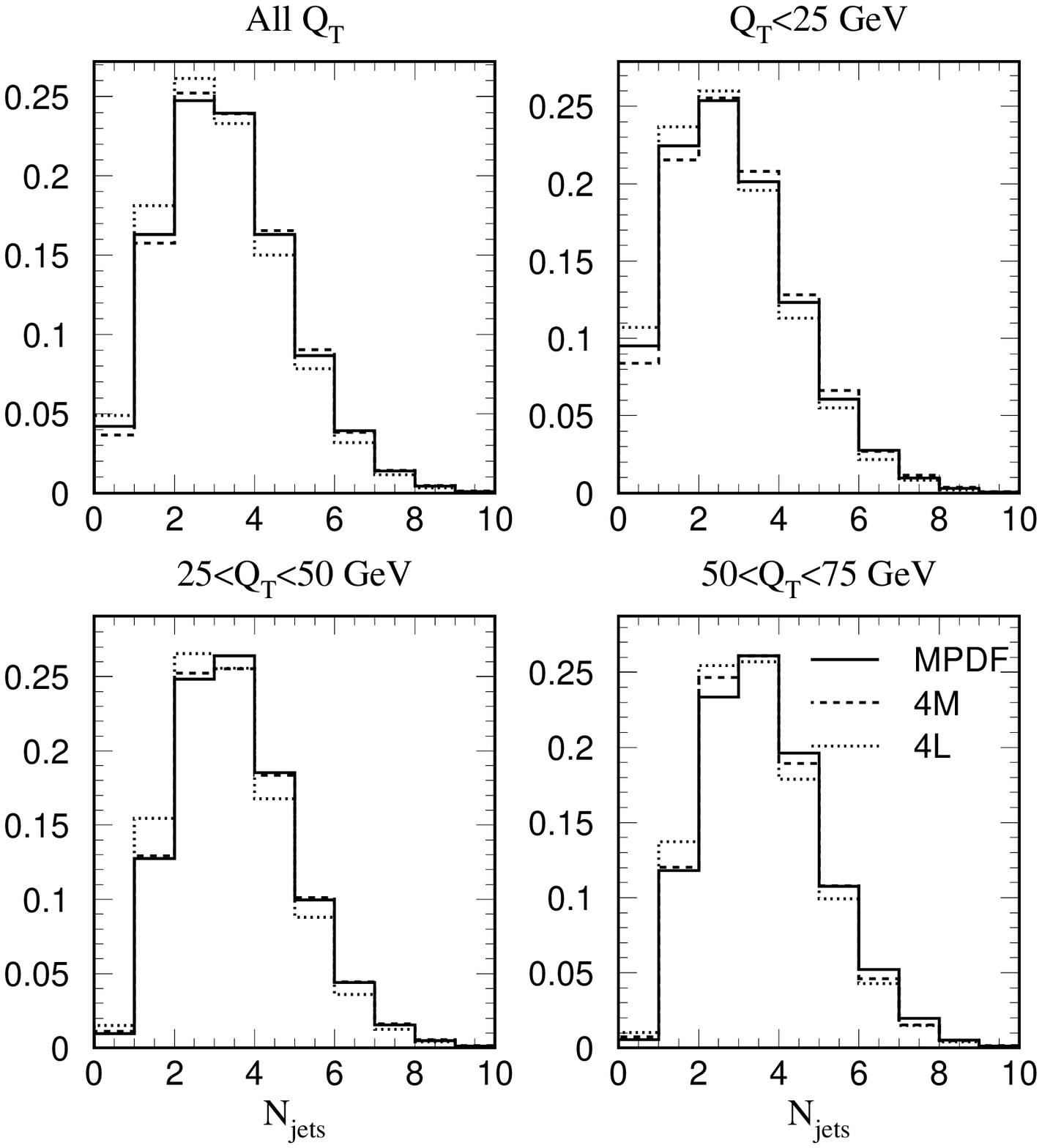,width=15cm}
\caption[]
{Jet multiplicity in Higgs boson events for different $Q_T$ binnings.}
\label{fig:hjets}
\end{figure}
\end{center}

\begin{center}
\begin{figure}
\psfig{figure=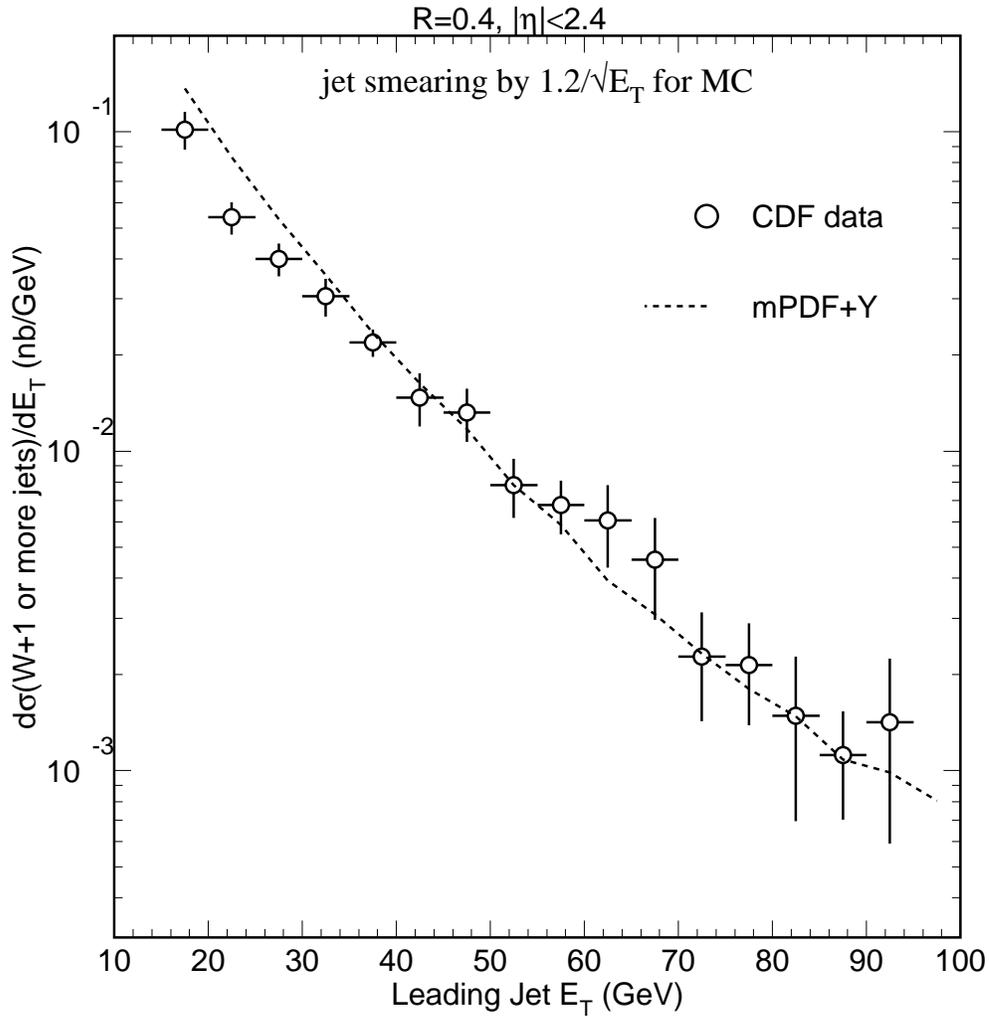,width=15cm}
\caption[]
{Comparison of the differential jet cross section in $W$ boson + jet events
in Run I (dashed line) to the CDF data (circles).   The jets have been
smeared by the resolution $\Delta E_T/E_T = 1.2/\sqrt{E_T}$ (in GeV).} 
\label{fig:dsdet}
\end{figure}
\end{center}

\end{document}